\documentclass[aps,prd,twocolumn,superscriptaddress,floatfix,nofootinbib,showpacs]{revtex4-1}
\usepackage{amsmath}

\usepackage{bm}
\usepackage{xcolor}
\usepackage{hyperref}
\usepackage{url}
\usepackage{graphicx}
\usepackage[large]{subfigure}
\usepackage{amssymb}
\usepackage[amssymb]{SIunits}

\usepackage{natbib}
\usepackage{multirow}
\usepackage{cleveref}

\newcommand{\apjl}{Astrophys J.}

\newcommand{\aap}{Astron. Astrophys.}

\newcommand{\mnras}{Mon. Not. R. Astron. Soc.}
\newcommand{\physrep}{Phys. Rep.}

\newcommand{\araa}{Ann.Rev.Astron.Astrophys.}

\newcommand{\bn}{{\bf n}}

\newcommand{\xinl}{\xi_{\rm{NL}}}
\newcommand{\tj}[6]{ \begin{pmatrix}
   #1 & #2 & #3 \\
   #4 & #5 & #6 
  \end{pmatrix}}

  \def\hp{ {\sc HEALPix}}

\begin{document}
\title{An excess of non-Gaussian fluctuations in the cosmic infrared background consistent with gravitational lensing}

\author{Chang Feng}
\email{changf@illinois.edu}
\affiliation{Department of Physics, University of Illinois at Urbana-Champaign, 1110 West Green Street, Urbana, Illinois, 61801, USA}

\author{Gilbert Holder}
\affiliation{Department of Physics, University of Illinois at Urbana-Champaign, 1110 West Green Street, Urbana, Illinois, 61801, USA}
\affiliation{Department of Astronomy, University of Illinois at Urbana-Champaign, 1002 West Green Street, Urbana, Illinois, 61801, USA}
\affiliation{Canadian Institute for Advanced Research, Toronto, Ontario M5G 1M1, Canada}

\begin{abstract}
The cosmic infrared background (CIB) is gravitationally lensed. A quadratic-estimator technique that is inherited from lensing analyses of the cosmic microwave background (CMB) can be applied to detect the CIB lensing effects. However, the CIB fluctuations are intrinsically strongly non-Gaussian, making CIB lensing reconstruction highly biased. We perform numerical simulations to estimate the intrinsic non-Gaussianity and establish a cross-correlation approach to precisely extract the CIB lensing signal from raw data. We apply this technique to CIB data from the {\em Planck} satellite and cross-correlate the resulting lensing estimate with the CIB data, galaxy number counts and the CMB lensing potential. We detect an excess that is consistent with a lensing contribution at $>4\sigma$. 
\end{abstract}

\maketitle

\section{Introduction}
\label{intro}
The cosmic infrared background (CIB) refers to the cumulative unresolved emission from dusty star-forming galaxies~\cite{1996A&A...308L...5P,2001ARA&A..39..249H}. As the CIB photons propagate across the universe, their trajectories are gravitationally distorted\cite{2018PhRvD..97l3539S}, just as do the photons of the cosmic microwave background (CMB)~\cite{2006PhR...429....1L}. Gravitational lensing effects on the CMB can leave direct observable signatures in both the temperature and polarization fluctuations. CMB lensing has been detected by both space~\cite{2014A&A...571A..17P, 2018arXiv180706210P} and ground-based experiments~\cite{2011PhRvL.107b1301D, 2012ApJ...756..142V, 2014PhRvL.113b1301A, 2015ApJ...810...50S, 2017PhRvD..95l3529S, 2016ApJ...833..228B}, but so far there is no detection of CIB lensing effects.

CIB lensing contains information about the redshift distribution of infrared sources, including the high-redshift sources that are difficult to detect at other wavelengths. The redshift distribution of infrared sources is closely related to the star formation history. With CIB fluctuations mapped at arc-second resolution from ongoing or future experiments, CIB lensing can probe fluctuations in the gravitational lensing potentials on smaller scales than CMB lensing, where the characteristic fluctuations are strongly suppressed on small scales. Fine-scale structure in the lensing potential can be useful for studying sterile neutrinos and even fuzzy dark matter~\cite{2019PhRvD..99b3502N}. Moreover, the matter distribution information from these small scales will shed light on baryonic physics, which could be used as a calibration tool for numerical simulations focusing on the impact of feedback and other astrophysical processes~\cite{2012ApJ...758...75B}. 
  
Although similarities exist between CMB and CIB fluctuations, the CIB has unique features making it more complicated than the CMB. Its emission spans a wide range of redshifts, enabling low redshift CIB sources to gravitationally distort infrared emission from high redshift ones~\cite{2004NewA....9..173C}. Another distinct feature of the CIB is that it is intrinsically non-Gaussian in the absence of the lensing effects.

In analogy to CMB lensing, squeezed triangles in the Fourier domain can be constructed to extract lensing information from CIB fluctuations. This approach is the so-called quadratic-estimator technique~\cite{2003PhRvD..67h3002O}. Application of a quadratic estimator to CIB fluctuations was theoretically investigated in~\cite{2018PhRvD..97l3539S} where it was found that the intrinsic non-Gaussian structure will cause additional noise and bias in auto correlations of the CIB lensing reconstructions. Cross-correlation techniques have proven to be an efficient method to mitigate various biases in CMB lensing studies~\cite{2004PhRvD..70j3501H,2007PhRvD..76d3510S,2008PhRvD..78d3520H,2012PhRvD..86f3519F}. However, as an outcome of the intrinsic non-Gaussianity induced by the non-linear growth of structure, the cross correlations between CIB lensing reconstructions and large-scale structure tracers also become complicated, because these cross correlations are non-vanishing even in the absence of the true lensing distortions. Contributions from the intrinsic non-Gaussianity must be removed in order to measure the CIB lensing effects.

In this work, we develop a cross-correlation technique to measure CIB lensing and intrinsic non-Gaussianity simultaneously. This paper is structured as follows: in Sec. \ref{theory}, we describe theoretical modeling of the cross correlations; in Sec. \ref{sim}, we discuss simulation details for the CIB lensing analysis; in Sec. \ref{validation}, we focus on calculations of the intrinsic non-Gaussianity at Planck high frequencies, then we apply the cross-correlation technique to the Planck CIB data in Sec. \ref{data} and conclude in Sec. \ref{con}.\\

\section{Clustering power spectrum}
\label{theory}
In this section we calculate various clustering power spectra using linear models. Three types of fluctuations---weak lensing convergence $\kappa$, CIB intensity $T$ and galaxy density contrast $g$---are considered in this work. A projected map for each of these fluctuations is a tracer of the matter distribution $\delta_m$, and can be uniformly expressed as
\begin{equation}
\Psi(\bn)=\int d\chi W^{\Psi}(\chi)\delta_m(d_A\bn), 
\end{equation} where $\Psi=\{\kappa, T, g\}$. $W^{\Psi}$ expresses the weight along the line of sight, $d_A$ is the angular diameter distance to a point at comoving distance $\chi$, and $\bn$ is a direction in the sky.

For weak lensing convergence, the lensing efficiency $W^{\kappa}$ for a source distribution $W^{s}$ is given by 
\begin{equation}
W^{\kappa}(\chi)=\frac{3\Omega_mH_0^2}{2} \frac{d_A}{a}\int_{\chi}d\chi'W^{s}(\chi')\frac{d_A(\chi'-\chi)}{d_A(\chi')},
\end{equation}
where angular diameter $d_A=\chi$ for a flat Universe, and $a$ is the cosmological scale factor. Cosmological parameters $\Omega_m$ and $H_0$ are the matter fraction and Hubble constant today. For CMB lensing, a single source distribution can be approximated as $W^{s}(\chi)=\delta(\chi-\chi_{\ast})$ where $\chi_{\ast}$ is the comoving distance to the last scattering surface. The lensing convergence is related to the potential via $\kappa=-(1/2)\nabla^2\phi$.

For the fluctuations of the CIB and galaxy density contrast, astrophysical information is encoded in galaxy biases $b$ which are assumed to be a simple single multiplicative factor (so-called ``linear bias'') such that  $W^{i}=b_{i}dN/dz$, where $i=\{T, g\}$.
In this work, we adopt the redshift distributions of Herschel at multiple frequencies~\cite{2013ApJ...771L..16H} to model the Planck CIB spatial clustering, and a galaxy redshift distribution model~\cite{2012MNRAS.422L..77G} $dN/dz\sim z e^{-(z-z_0)^2/2\sigma^2_z}$ with $z_0=0.1$ and $\sigma_z=0.1$, for the galaxy map generated from the Wide-field Infrared Survey Explorer (WISE). Galaxy biases for both data sets are determined from Planck CIB auto-power spectra $\langle TT\rangle$ ($\sim b^2_T$) and cross-power spectra between Planck lensing and the WISE data $\langle \phi g\rangle$ ($\sim b_g$), respectively.

The clustering power spectrum is 
\begin{equation}
C^{\Psi\Psi'}_{\ell}=\int_0^{z_s} \frac{d\chi}{\chi^2}W^{\Psi}W^{\Psi'}P(k,z),
\end{equation}
where $P(k,z)$ is the matter power spectrum at spatial wavenumber $k$ and redshift $z$. In this work, a redshift cutoff is set to a sufficiently high value $z_s=100$ and the latest Planck cosmological parameters are used~\cite{2018arXiv180706209P}. In the rest of the text, $\kappa$ (or $\phi$) refers to the CIB lensing and $\kappa^{\rm CMB}$ (or $\phi^{\rm CMB}$) refers to the CMB lensing.

\section{CIB lensing simulations and reconstructions}
\label{sim}
Analogous to CMB lensing, the auto-power spectrum $\langle \phi\phi\rangle$ of a CIB lensing potential $\phi$ could be used to probe the lensing effects in the Planck CIB data, but the expected detection significance is only at the $\sim1\sigma$ level due to insufficient sensitivity of the Planck high frequency data~\cite{2018PhRvD..97l3539S}. Cross power spectra can improve the detection significance of a noisy signal and in this analysis we will focus on the cross-power spectra $\langle\phi T\rangle$, $\langle\phi g\rangle$, and $\langle\phi\phi^{\rm CMB}\rangle$.

The tracers $\Psi$ are assumed to be Gaussian fluctuations which can be directly generated from their power spectra. However, these tracers are also correlated so a Cholesky decomposition of the covariance, built on 21 auto- and cross-power spectra among six fields---$\phi^{\rm CMB}$, $\phi (857{\rm GHz})$, $\phi (545{\rm GHz})$, $T(857{\rm GHz})$, $T (545{\rm GHz})$ and $g$---is performed to make these fields correlated. With the decomposition, these fields are expressed as
\begin{equation}
\begin{bmatrix}
\phi^{\rm CMB}_{\ell m}\\
\phi^{\alpha}_{\ell m} \\
\phi^{\beta}_{\ell m} \\
T^{\alpha}_{\ell m}\\
T^{\beta}_{\ell m}\\
g_{\ell m}
\end{bmatrix}
=\begin{bmatrix}
f^{{\phi}^{\rm CMB}}_{\ell}&&&\\
f^{{\phi^{\alpha} }}_{1,\ell}&f^{{\phi^{\alpha}}}_{2,\ell}&&\\
f^{{\phi^{\beta}}}_{1,\ell}&f^{{\phi^{\beta}}}_{2,\ell}&f^{{\phi^{\beta}}}_{3,\ell}&\\
f^{{T^{\alpha}}}_{1,\ell}&f^{{T^{\alpha}}}_{2,\ell}&f^{{T^{\alpha}}}_{3,\ell}&f^{{T^{\alpha}}}_{4,\ell}\\
f^{{T^{\beta}}}_{1,\ell}&f^{{T^{\beta}}}_{2,\ell}&f^{{T^{\beta}}}_{3,\ell}&f^{{T^{\beta}}}_{4,\ell}&f^{{T^{\beta}}}_{5,\ell}\\
f^{g}_{1,\ell}&f^{g}_{2,\ell}&f^{g}_{3,\ell}&f^{g}_{4,\ell}&f^{g}_{5,\ell}&f^{g}_{6,\ell}
\end{bmatrix}
\begin{bmatrix}
G^{{\phi}^{\rm CMB}}_{\ell m}\\
G^{\phi,\alpha}_{\ell m}\\
G^{\phi,\beta}_{\ell m}\\
G^{T^{\alpha}}_{\ell m}\\
G^{T^{\beta}}_{\ell m}\\
G^g_{\ell m}
\end{bmatrix}\label{corrsims}.
\end{equation}
Here $G_{\ell m}$s are Gaussian random variables drawn from a normal distribution, and $f_{\ell}$s are derived from the theoretical covariance matrix. CIB maps at two different frequencies $\alpha$ and $\beta$ are considered for this analysis, and $\alpha$ and $\beta$ refer to 857 GHz and 545 GHz, respectively. We verified that the six maps simulated by this decomposition, can reproduce all the 21 auto- and cross-power spectra as the input ones. We further apply the lensing procedure $T^{\nu}({\textbf n}+\nabla \phi^{\nu})$ to make the CIB fields lensed at both frequencies $\nu$ using the software \textit{Taylens}~\cite{2013MNRAS.435.2040L}.

For a CIB field $T$ with an intrinsic non-Gaussianity, we assume that we can capture this non-Gaussianity with a local model where the non-Gaussian field is simply a function of an underlying Gaussian field $T_0$, i.e., $T=f(T_0)$. The non-Gaussian CIB field can be Taylor-expanded, and for this study we keep only the first order term, i.e.,
\begin{equation}
T=T_0+\xinl T^2_0\label{toy},
\end{equation}
so the level of non-Gaussianity is quantified by a single parameter, in analogy to the primordial non-Gaussianity problem~\cite{2016A&A...594A..17P, 2015PhRvD..92d3509F}.
With this quadratic term, CIB bispectra $\langle TTT\rangle$ and $\langle TTg\rangle$ become non-zero even if there is no lensing signal in the CIB fluctuations. Thus, the model in Eq. (\ref{toy}) provides a way to simulate the intrinsic non-Gaussian fluctuations, i.e., $\Delta T^{\rm NG}=\xinl T^2_0$. Below, we will verify that this local non-Gaussian model captures much of the complexity of the Planck CIB data.

In addition to the correlated Gaussian simulations that only contain signal pieces, Gaussian noise maps for $\Psi$ are also created from their power spectra, which are Planck lensing-reconstruction noise, noise power spectra that combine both residual instrumental noise and residual dust for the CIB component-separated maps, and a WISE-like shot noise, corresponding to fields $\phi$, $T$ and $g$, respectively. Experimental specifications are also incorporated into the CIB simulations, including a $\theta=5'$  Gaussian beam and a conservative sky cut.

\begin{figure*}
\includegraphics[width=8cm, height=8cm]{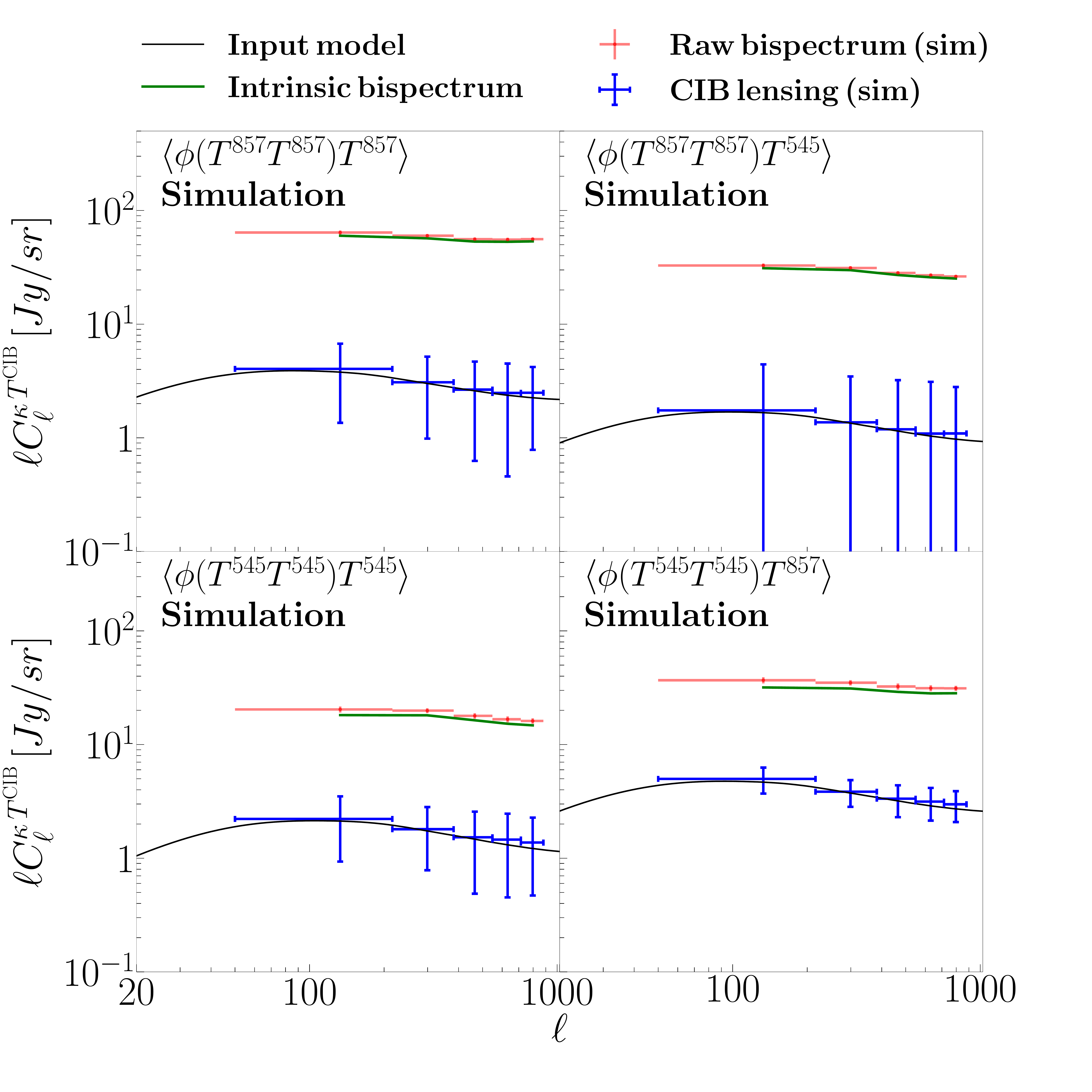}
\includegraphics[width=8cm, height=8cm]{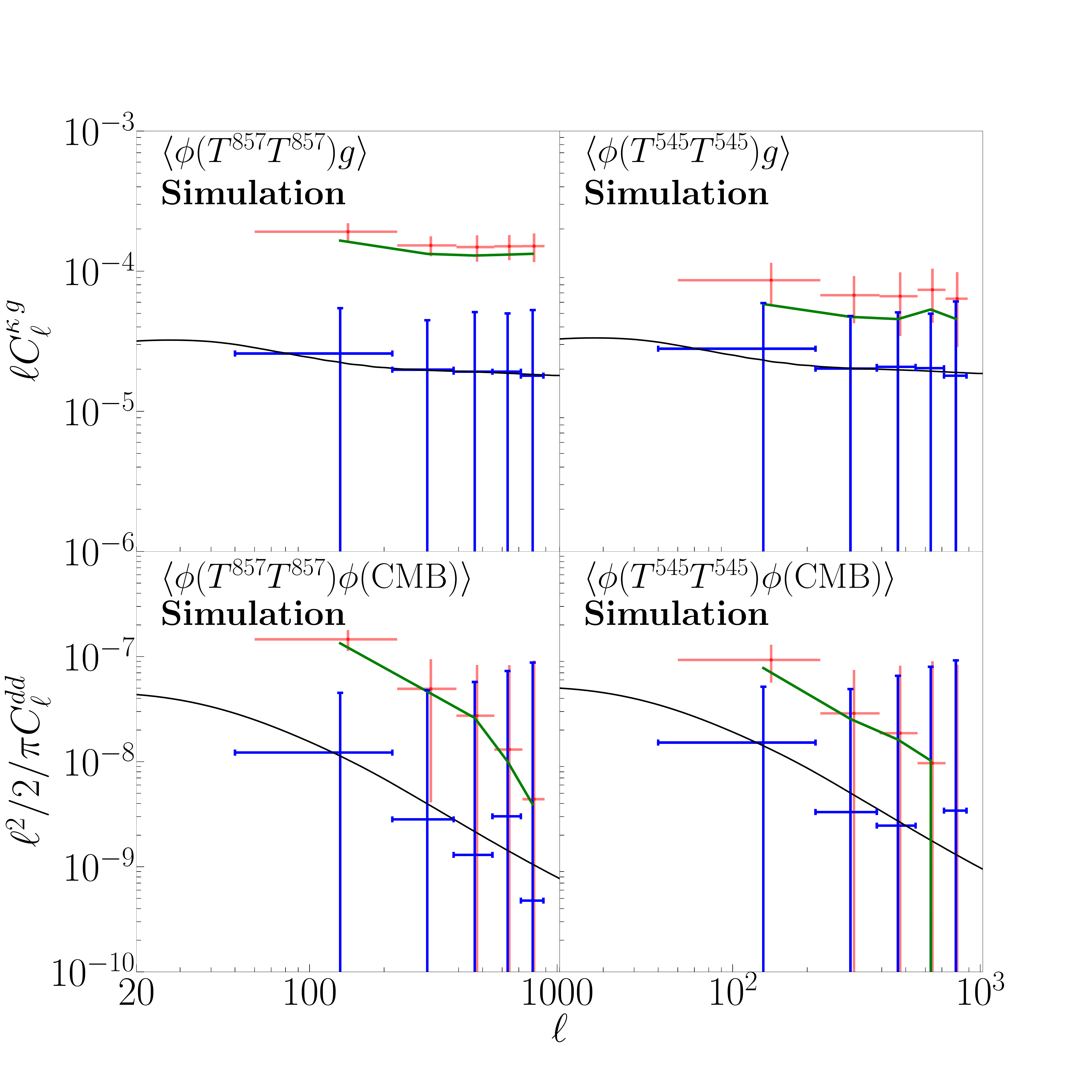}
\caption{Validation of all the cross correlations $\langle\phi T\rangle$, $\langle\phi g\rangle$ and $\langle\phi\phi^{\rm CMB}\rangle$. Even through the intrinsic bispectra (green lines) are roughly one order of magnitude higher than the CIB lensing signals (black lines), the latter can be correctly separated from the raw cross correlations (red points) between CIB lensing reconstructions and tracers. The unbiased cross-power spectra of the lensing signals are shown with blue data points. The band powers of the raw bispectra in the right panel are shifted for clarity. For the $y$-axis of the bottom figure in the right panel, a power spectrum of deflection field $d$ which is related to the lensing potential via $d=\nabla\phi$ is shown. This convention applies to the other figures unless otherwise noted.}\label{valid_kt}
\end{figure*}

The CIB simulations produced in the aforementioned steps are dedicated to validation purposes and are used to estimate different bispectra and biases involved in data analysis. Two sets of CIB simulations $T^{(0)}$ and $T^{(f)}$ are made and they are defined as ``unlensed $+$ non-Gaussian'' CIB maps
\begin{equation}
T^{(0)}=T_0+\xinl T_0^2\label{t0}
\end{equation}
and ``lensed $+$ non-Gaussian'' maps
\begin{equation}
T^{(f)}=T_0(\bn+\nabla\phi)+\xinl T_0^2.\label{tf}
\end{equation}
Here $T_0$ contains neither lensing nor non-Gaussian effects and the parameter $\xinl$ has units of one over the mean intensity of the CIB. In principle, the non-Gaussianity in the CIB is itself lensed, but we leave that for future work. The CIB lensing maps $\hat\phi(\bn)$ can be reconstructed from the ``lensed $+$ non-Gaussian'' maps following a standard quadratic estimator~\cite{2003PhRvD..67h3002O}, i.e., 
\begin{equation}
\hat\phi(\bn)\sim\nabla^i[A(\bn)\nabla_i B(\bn)],\label{qe}
\end{equation}
where the filtered maps
\begin{equation}
A(\bn)=\sum_{\ell m}\frac{1}{C_{\ell}}T_{\ell m}Y_{\ell m}(\bn)
\end{equation}
and
\begin{equation}
B(\bn)=\sum_{\ell m}\frac{\tilde C_{\ell }}{C_{\ell}}T_{\ell m}Y_{\ell m}(\bn)
\end{equation}
are each created for the two types of CIB simulations. Here $\nabla$ is the covariant gradient on a unit sphere, $\tilde C_{\ell}$ and $C_{\ell}$ are unlensed and observed CIB power spectra. Spherical harmonic modes $T_{\ell m}$s are transformed from the CIB maps, and $Y_{\ell m}$ is a spin-0 spherical harmonic. 

The CIB lensing reconstruction is dominated by high-$\ell$ signals for Planck so the CIB multipole range is chosen as $1024<\ell<2048$, which can also conservatively remove the large-scale Galactic dust contamination~\cite{2018PhRvD..97l3539S}. For the tracers $\Psi$, the multipole range used is $2<\ell<1024$. The noisy reconstruction described in Eq. (\ref{qe})~\cite{2003PhRvD..67h3002O} essentially forms a triangle in harmonic space with two legs being CMB modes $X_{\ell m}$ and $Z_{\ell'm'}$. An equivalent form to the real-space definition Eq.(\ref{qe}), can be expressed in the Fourier domain as
\begin{eqnarray}
\hat\phi_{LM}(X_{\ell m},Z_{\ell'm'})&=&A_L\displaystyle\sum_{\ell m\ell'm'}(-1)^M\tj{\ell}{\ell'}{L}{m}{m'}{-M}\nonumber\\
&\times&g_{\ell\ell'}(L)X_{\ell m}Z_{\ell'm'},\label{qe1}
\end{eqnarray}
where the big bracket $(...)$ is the 3-$j$ Wigner symbol. The normalization function $A_L$ and weighting function $g_{\ell\ell'}(L)$ are given in ~\cite{2003PhRvD..67h3002O}. 

We run lensing reconstructions and obtain two sets of $\phi$ maps---$\hat\phi^{(0)}$ and $\hat\phi^{(f)}$, corresponding to the two types of CIB simulations $T^{(0)}$ and $T^{(f)}$, respectively. With these simulations, an estimator of the cross correlation between CIB lensing and the tracers is established for a particular choice of $\xinl$ 
\begin{equation}
{C}_{\ell}^{\hat\phi\Psi}(\xinl)=\langle\hat\phi^{(f)}\Psi\rangle-\langle\hat\phi^{(0)}\Psi\rangle,\label{rawest}
\end{equation}
where $\langle\hat\phi^{(f)}\Psi\rangle$ and $\langle\hat\phi^{(0)}\Psi\rangle$ are referred to as the raw and intrinsic bispectra, respectively, and $\hat \phi$ denotes the biased lensing reconstruction. With the definitions of the CIB maps defined in Eqs.(\ref{t0}) and (\ref{tf}), the bispectra in Eq. (\ref{rawest}) can be expanded into a series of correlation functions where only Gaussian fields $T_0$ and $\phi$ are involved. Beyond the leading terms such as the four-point correlations $\mathcal{O}(\phi T_0^3)$ and $\mathcal{O}(T_0^4)$ in the raw bispectra, higher order correlations are also present. Gravitational lensing effects can introduce secondary CIB fluctuations $\Delta T^{\phi,(n)}\sim\phi^{n} T_0$, where the power $n$ is the order in $\phi^{n}$. The first order perturbation $\Delta T^{\phi,(1)}\sim\phi T_0$ is the lensing signal reconstructed by the estimator. Higher order perturbations such as $\Delta T^{\phi,(2)}\sim\phi^2 T_0$ are coupled to the intrinsic non-Gaussian fluctuations $\Delta T^{\rm NG}=\xinl T_0^2$, giving rise to a non-vanishing bispectrum $\langle T_0\Delta T^{\phi,(2)}\Delta T^{\rm NG}\rangle$ which contains six-point correlation functions like $\mathcal{O}(\phi^3T_0^3)$ and $\mathcal{O}(\phi^2T_0^4)$. If unaccounted for, this type of bispectrum can bias the estimation. Using the estimator (Eq. (\ref{rawest})), this bias is estimated as
\begin{equation}
\Delta C^{\rm higher}_{\ell}={C}_{\ell}^{\hat\phi\Psi}|_{\xinl=1}-{C}_{\ell}^{\hat\phi\Psi}|_{\xinl=0}.\label{higher}
\end{equation}
A debiased estimation of the desired cross correlation $\langle\phi\Psi\rangle$ is expressed as
\begin{equation}
{C}_{\ell}^{\phi\Psi}={C}_{\ell}^{\hat\phi^{(f)}\Psi}-{C}_{\ell}^{\hat\phi^{(0)}\Psi}(\xinl)-\xinl\Delta C^{\rm higher}_{\ell}\label{unbiased},
\end{equation}

or alternatively, the raw bispectrum is
\begin{equation}
{C}_{\ell}^{\hat\phi^{(f)}\Psi}={C}_{\ell}^{\phi\Psi}+{C}_{\ell}^{\hat\phi^{(0)}\Psi}(\xinl)+\xinl\Delta C^{\rm higher}_{\ell}\label{unbiased_raw}.
\end{equation}For the measured raw bispectrum, $\hat\phi^{(f)}$ is replaced by the lensing reconstruction from data.

With this estimator, three types of cross-power spectra $\langle\phi T\rangle$, $\langle\phi g\rangle$, and $\langle\phi\phi^{\rm CMB}\rangle$ can be measured. At a single frequency, these measurements are combined and a joint measurement is constructed through
\begin{equation}
{\bf x}^D=({\bf C}^{\phi T}, {\bf C}^{\phi g}, {\bf C}^{\phi \phi^{\rm CMB}}).\label{cov}
\end{equation}Here $\bf x$ and $\bf C$ are vectors which align the power spectra values according to the sequence of $\ell$s. The covariance matrix for the joint measurement
\begin{equation}
\mathcal{C}=\langle(\bf x -\bf \bar x)(\bf x -\bf \bar x)\rangle
 \end{equation}
 is derived from an ensemble of simulations for the CIB lensing reconstructions and the tracers. As the structure of the raw bispectrum in Eq. (\ref{unbiased_raw}), a model of the joint measurement is assumed as 
 \begin{equation}
 {\bf x}=A^{\rm CIB}_{\rm lens}{\bf C}+{\tilde\xi}_{\rm NL} [{\bf {\tilde C}}^{\rm intrinsic}+\Delta {\bf C}^{\rm higher}],\label{model}
 \end{equation}where ${\bf {\tilde C}}$ and $\Delta {\bf C}^{\rm higher}$ are both calculated at ${\xinl}_{\ast}=2$. The parameter $A^{\rm CIB}_{\rm lens}$ is an overall amplitude ratio of measured CIB lensing power-spectrum to the theoretical expectation ${\bf C}^{(\lambda)}$ and a reduced $\tilde\xi_{\rm NL}$ is defined as $\tilde\xi_{\rm NL}=\xinl/{\xinl}_{\ast}$. With these definitions, a maximum likelihood can be established as 
 \begin{equation}
 -2\ln\mathcal{L}=(\bf x^D-\bf x)\mathcal{C}^{-1}(\bf x^D-\bf x).\label{like}
 \end{equation}Fig. \ref{valid_kt} shows that all the cross-power spectra are dominated by the intrinsic bispectra and not the lensing signals at ${\xinl}_{\ast}=2$. Even though the lensing signal is about an order of magnitude fainter than the intrinsic ones, the unbiased estimator, described in Eq. \ref{unbiased}, can precisely extract the cross-power spectra $\langle\phi T\rangle$, $\langle\phi g\rangle$ and $\langle\phi\phi^{\rm CMB}\rangle$ from the CIB simulations with no significant biases.

\begin{figure}
\includegraphics[width=8cm, height=5cm]{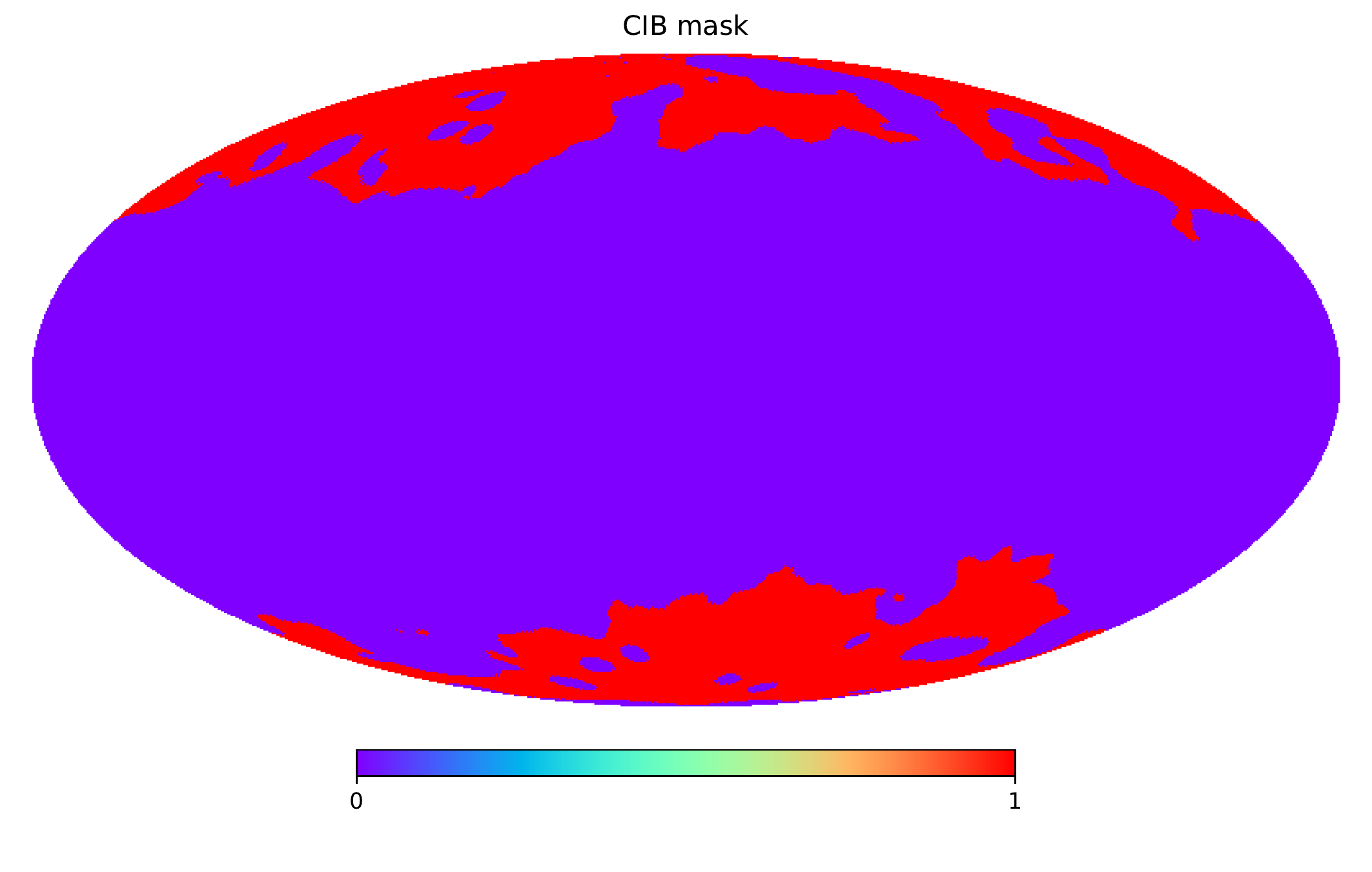}
\caption{A mask of Planck CIB data is formed from CIB component separation scheme GNILC, and the sky coverage is 17\%. The GNILC CIB map is almost dust-free in that high-latitude regions are optimally selected so the CIB reconstruction is less contaminated by Galactic dust.}\label{mask}
\end{figure}

\begin{figure}
\includegraphics[width=9cm, height=9cm]{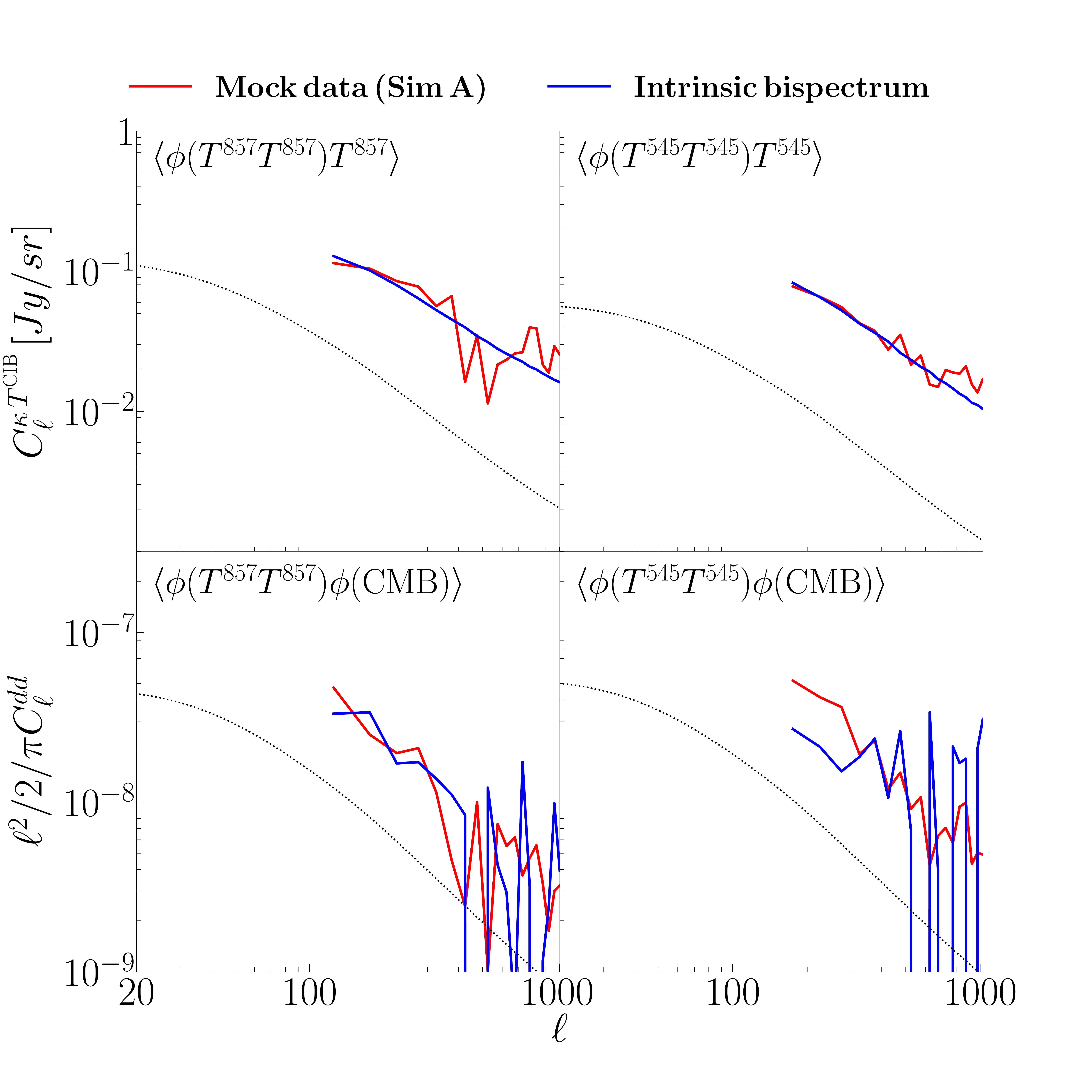}
\caption{Validations for two cross-power spectra $\langle\phi T\rangle$ (top) and $\langle\phi\phi^{\rm CMB}\rangle$ (bottom) with \textbf{Sim A} at 857 GHz and 545 GHz. The residual bispectra are consistent with zero. \textbf{Sim A} only made one realization for the CIB map $T$ at each frequency and one realization for the CMB lensing potential $\phi^{\rm CMB}$ so the two cross-power spectra are formed. For reference, the input models for CIB lensing at 857 and 545 GHz are shown in black dotted curves.}\label{valid_toronto_high}
\end{figure}

\begin{figure}
\includegraphics[width=9cm, height=9cm]{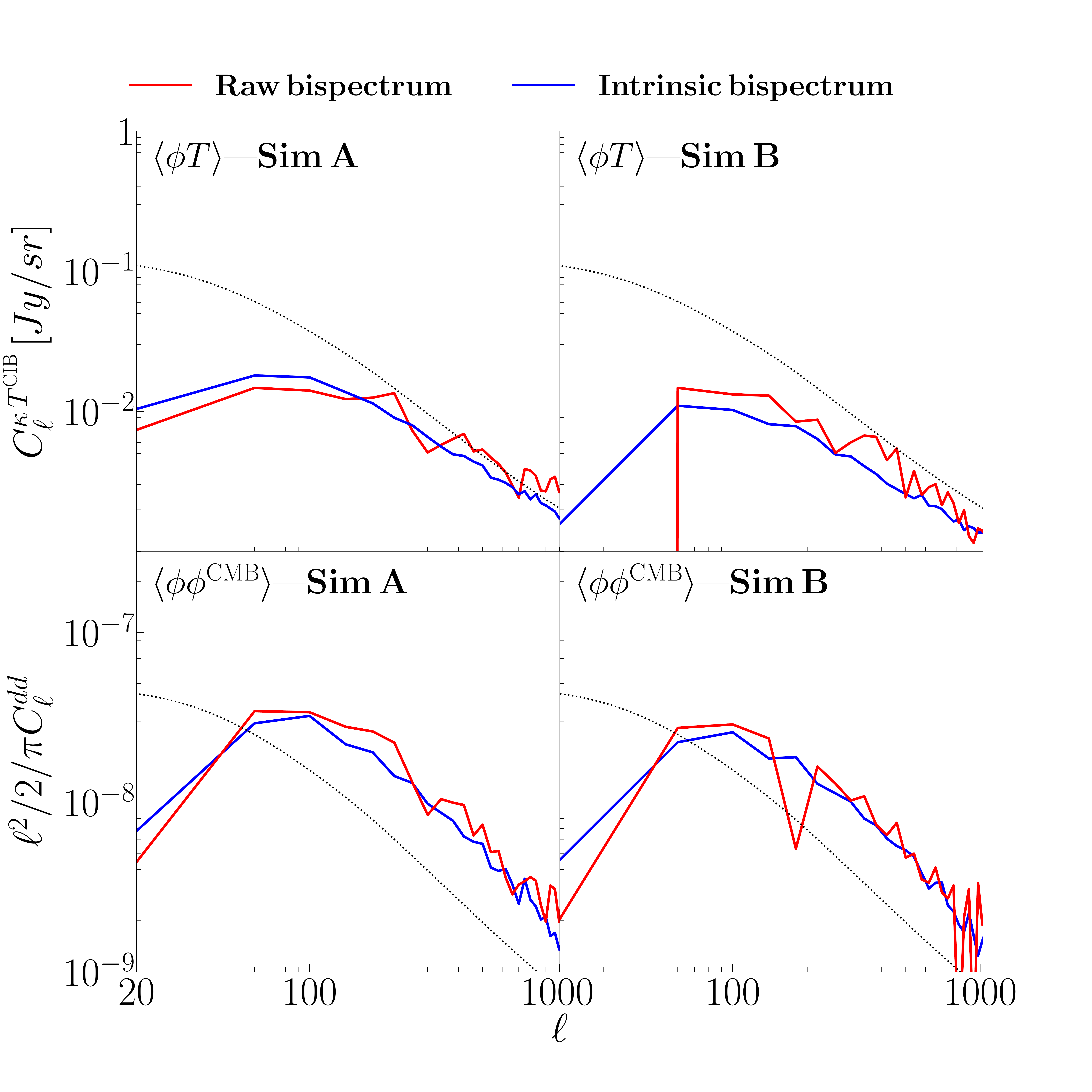}
\caption{Validation for two cross-power spectra $\langle\phi T\rangle$ (top) and $\langle\phi\phi^{\rm CMB}\rangle$ (bottom) with two sets of mock simulations at 353 GHz---\textbf{Sim A}  (left) and \textbf{Sim B}  (right). The figure shows that the raw bispectra produced by \textbf{Sim A}  (left) and \textbf{Sim B} are consistent. For reference, the theoretical predictions at 857 GHz are shown in black dotted curves.}\label{valid_toronto_353}
\end{figure}

\begin{figure}
\includegraphics[width=9cm, height=9cm]{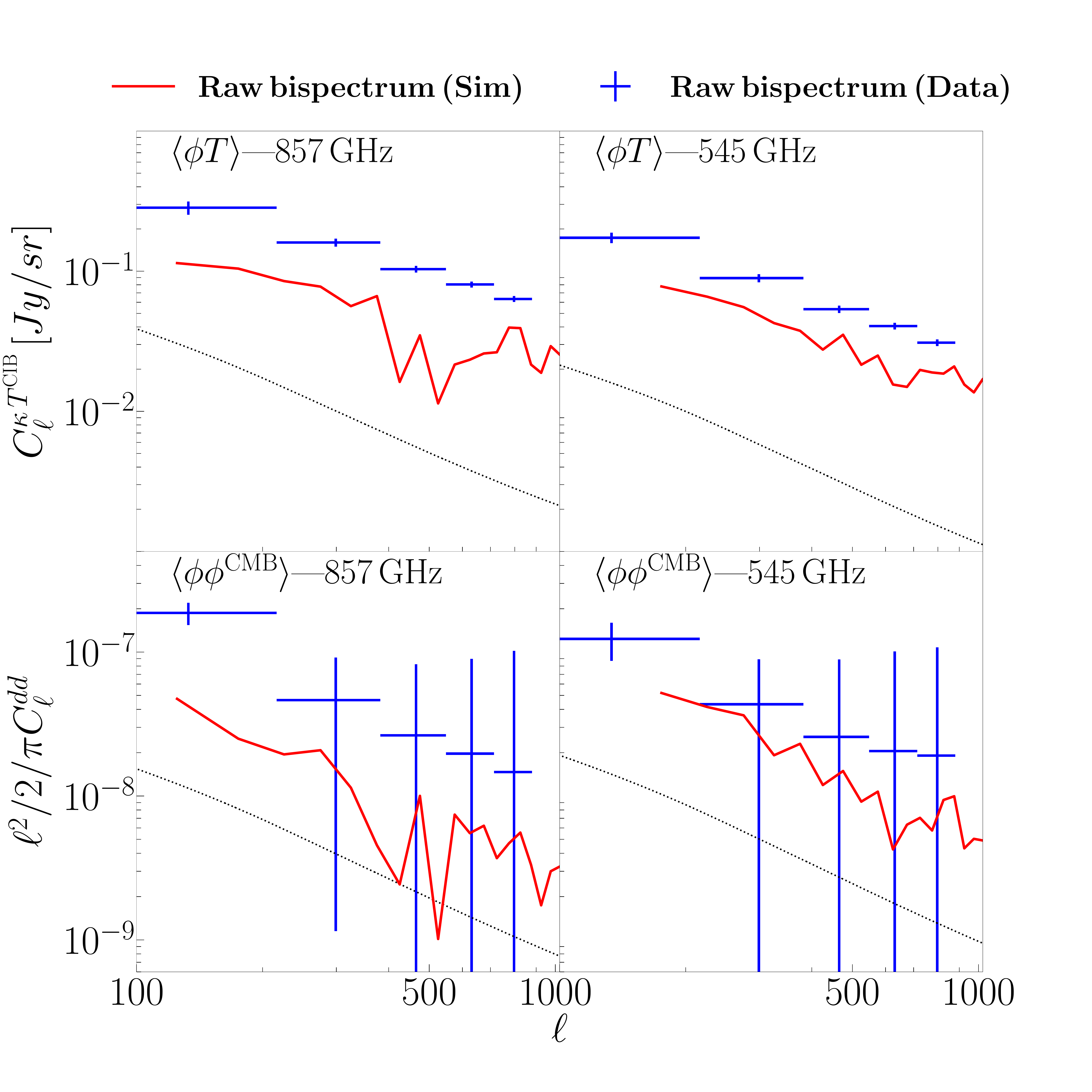}
\caption{Consistency check of the raw bispectra $\langle\phi T\rangle$ (top) and $\langle\phi\phi^{\rm CMB}\rangle$ (bottom) which are calculated from \textbf{Sim A} (red) and Planck data (blue). The black dotted curves are input models for CIB lensing. In each panel, the bispectrum shapes are consistent, although the amplitude of the simulations are well below that of the Planck data.}\label{mockanddata}
\end{figure}

\begin{figure*}
\includegraphics[width=16cm, height=16cm]{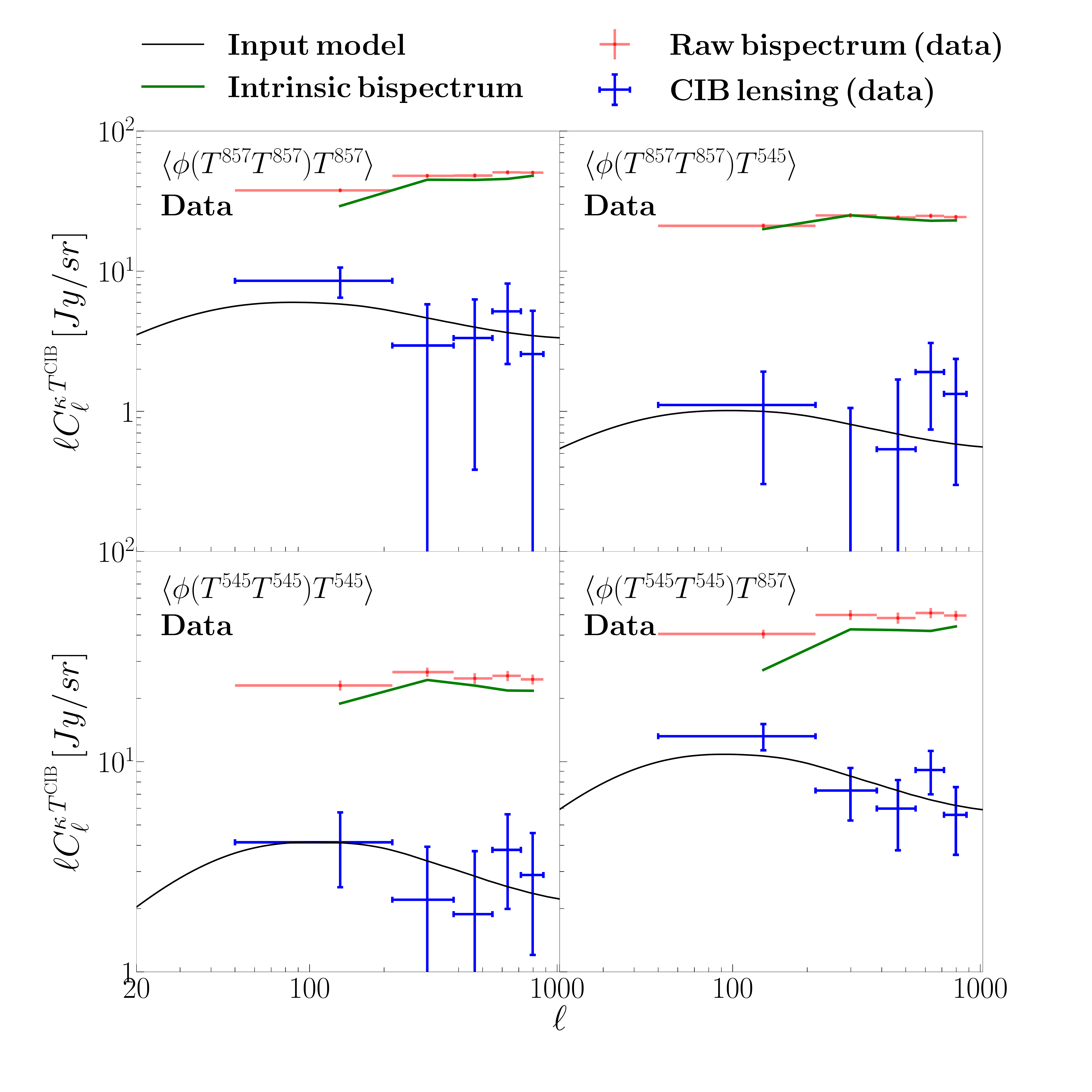}
\caption{Measurements of the cross correlations $\langle\phi(T,T)T'\rangle$ at Planck frequencies 857 and 545 GHz. Here $T$ or $T'$ refers to either 857 GHz or 545 GHz and a specific frequency permutation is labeled in each panel. The calculations confirm that the intrinsic bispectra, that arise from the intrinsic non-Gaussianity of the CIB, dominate the raw bispectra. The cross-correlation technique described in Sec. \ref{sim} can effectively separate the CIB lensing signals from the raw bispectra, and is applied to the Planck data at these two frequencies. The measured bispectrum residuals show an excess of the CIB non-Gaussianity, which is consistent with a CIB lensing signal.}\label{datakt}
\end{figure*}

\begin{figure*}
\includegraphics[width=16cm, height=16cm]{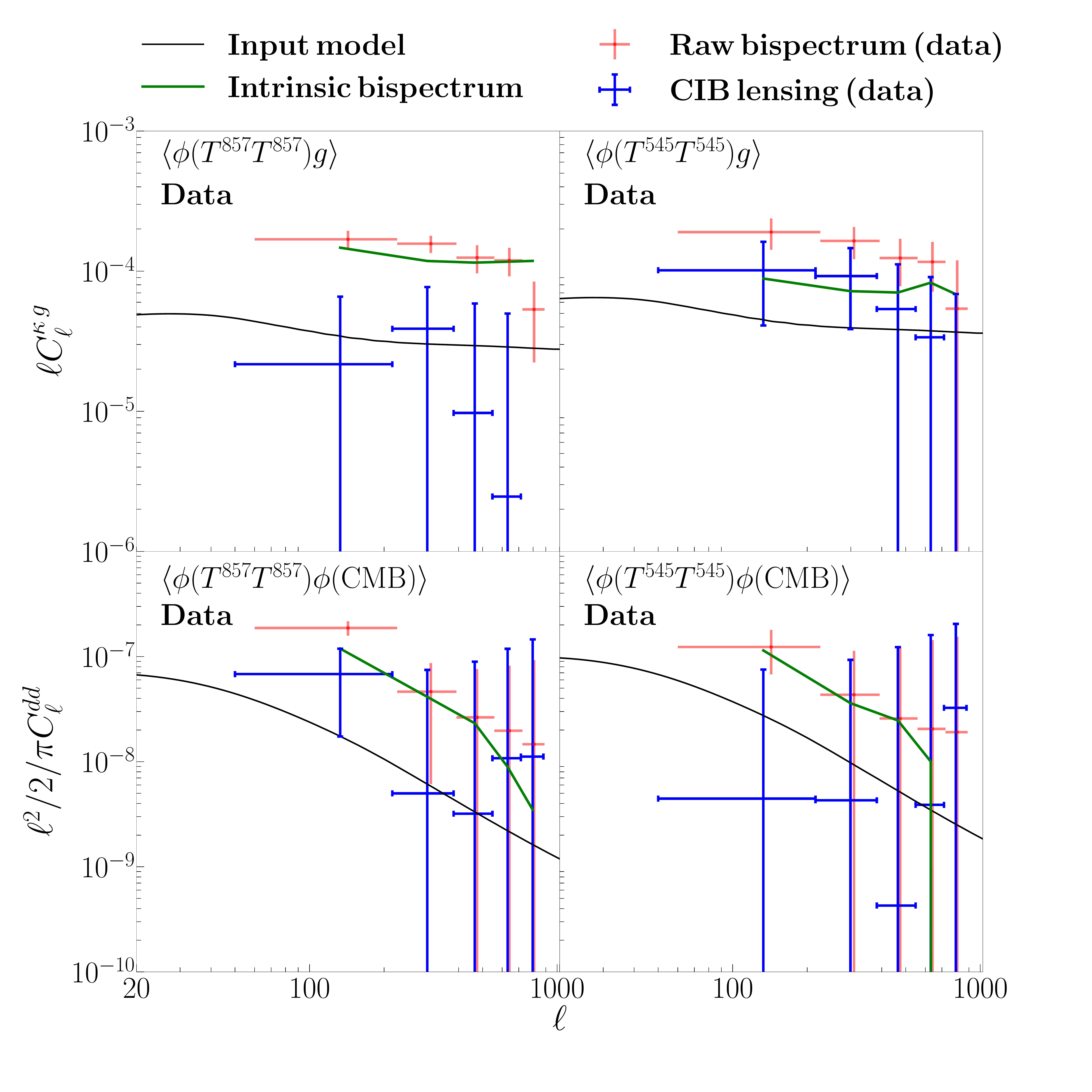}
\caption{Measurements of the cross correlations $\langle\phi(T,T)g\rangle$ and $\langle\phi(T,T)\phi^{\rm CMB}\rangle$ at Planck frequencies 857 and 545 GHz. Here $g$ is a map of galaxy number counts from the WISE catalog, and $\phi^{\rm CMB}$ is the minimum-variance lensing map from the Planck 2015 release 2. The band powers of the raw bispectra are shifted for clarity. Descriptions of the panels are the same as Fig. \ref{datakt}. The residual and intrinsic bispectra are separated from the joint frequency data $\langle\phi^{\nu} T^{\nu}\rangle+\langle\phi^{\nu} g\rangle+\langle\phi^{\nu} \phi^{\rm CMB}\rangle$ with $\nu=$ 857, 545 GHz. This gives rise to a slight offset between the raw and intrinsic bispectra in the top right panel at 545 GHz, because the cross correlations with tracers do not have enough sensitivity to constrain the model (Eq. (\ref{model})) due to the small sky coverage and the various tracer noises.}\label{datakgp}
\end{figure*}

\section{Validations of the local non-Gaussian model}
\label{validation}

The cross-correlation scheme described in Sec. \ref{sim} is built assuming a particular approximation to the form of the intrinsic non-Gaussianity. Before applying the technique to the data, the local model Eq. (\ref{toy}) must be checked to make sure it can produce consistent bispectrum shapes with CIB simulations generated from more realistic simulations. Two sets of simulations are investigated in this work. One set of CIB realizations (\textbf {Sim A}) is made from a 2LPT based cosmological simulation with $12288^3$ particles in a 15.4 Gpc box\footnotemark[1]\footnotetext[1]{\url{https://mocks.cita.utoronto.ca/index.php/WebSky_Extragalactic_CMB_Mocks}}. Based on a halo catalog, CIB realizations are produced at three Planck frequencies---353, 545 and 857 GHz~\cite{2019MNRAS.483.2236S}. Another CIB realization (\textbf{Sim B}) is described in~\cite{2010ApJ...709..920S}\footnotemark[2]\footnotetext[2]{\url{https://lambda.gsfc.nasa.gov/toolbox/tb_cmbsim_ov.cfm}}, where a Tree-Particle-Mesh N-body code is used to evolve dark matter particles in a simulation box. Ray-tracing through a population of dark matter halos, emission from infrared galaxies is mapped at 353 GHz. While we do not use 353 GHz Planck data in our analysis, \textbf{Sim B} serves as a check on variation in theoretical models for the CIB.

These CIB simulations trace emissions of infrared galaxies below $z\sim4$ and are quite non-Gaussian, as seen from the pixel histograms. We add correlated Gaussian modes to account for emissions above $z\sim4$. These CIB simulations are further synthesized in the same way as the Planck data: the instrumental noises are added, and a mask is applied (Fig. (\ref{mask})). \textbf{Sim A} made one realization for the CIB map $T$ at each frequency and one realization for the CMB lensing potential $\phi^{\rm CMB}$ so two cross-power spectra---$\langle\phi T\rangle$ and $\langle\phi\phi^{\rm CMB}\rangle$---are formed for testing the local model. We make three mock CIB maps at the three frequencies, and produce a suite of simulations for $T$ and $\phi^{\rm CMB}$ which are Gaussian and correlated, using the power spectra of the mock data. The error bars are derived from the simulation ensemble. From Fig. \ref{valid_toronto_high}, it is seen that the local model can well fit the raw bispectra of the mock data at 857 and 545 GHz, at which frequencies the Planck data are analyzed. Moreover, we estimate the residual bispectra after subtracting the intrinsic ones and find that they are consistent with zero at both frequencies as no CIB lensing signals are incorporated into the simulations. 

We make the mock data for both \textbf {Sim A} and \textbf {Sim B} at the frequency 353 GHz in a slightly different way, in which the mock data only consist of low-$z$ and high-$z$ signal pieces. No instrumental effects are incorporated for the mock data at this frequency, because we do not use the Planck 353 GHz data in this work. Tests at this frequency aim at checking consistency between the mock data and the local model. Again, Fig. \ref{valid_toronto_353} shows that the local model can produce sufficiently reasonable bispectrum shapes as the numerical simulations.

From Figs. \ref{valid_toronto_high} and \ref{valid_toronto_353}, one interesting feature of the $\langle\phi T\rangle$ bispectrum is that the amplitude of the intrinsic bispectrum decreases as the wavelength increases. This indicates that lower frequencies would be better suited for CIB lensing measurements because the intrinsic bispectra might become subdominant to the CIB lensing bispectra. However, at the lower frequencies, the CMB will come into play and the ability to detect CIB lensing would be greatly compromised.

As another sanity check of $\textbf{Sim A}$, we compute its raw bispectra and compare them with the Planck bispectra in Fig. \ref{mockanddata}. We omit the similar tests for $\textbf{Sim B}$ because CIB realizations are made up to the highest frequency 353 GHz and the Planck data at this frequency is not considered. Although there are some noticeable differences in the amplitudes as seen from the figure, the bispectrum shapes of the mock data are consistent with the Planck data. The models were not tuned to match the Planck data, suggesting that there is room for improvement in the detailed modeling of the CIB in the simulations. We defer a scrutiny of the discrepancy to future work. Nevertheless, these tests show that both the mock data and the local non-Gaussian model (Eq. (\ref{toy})) can produce consistent bispectrum shapes with the Planck data.

\begin{table*}
\small
\caption{Detection significance for measured bispectrum residuals. In the table, $\phi^{857}$ and $\phi^{545}$ refer to the CIB lensing maps reconstructed from Planck CIB data at 857 GHz and 545 GHz, respectively. A reduced $\tilde\xi_{\rm NL}$ is defined as $\tilde\xi_{\rm NL}=\xinl/{\xinl}_{\ast}$. Both parameters $A^{\rm CIB}_{\rm lens}$ and $\tilde\xi_{\rm NL}$ are marginalized values. The mean intensities $\bar I$ of the CIB at 857 and 545 GHz are 0.576 and 0.371 MJy/sr~\cite{2019arXiv190411556O}.}
\begin{center}{
\begin{tabular}{c|c|c|c|c} 
\hline
data&$\langle\phi^{857} T^{857}\rangle+\langle\phi^{857} g\rangle+\langle\phi^{857} \phi^{\rm CMB}\rangle$&$\langle\phi^{857} T^{545}\rangle$&$\langle\phi^{545} T^{545}\rangle+\langle\phi^{545} g\rangle+\langle\phi^{545} \phi^{\rm CMB}\rangle$&$\langle\phi^{545} T^{857}\rangle$\\
\hline
$A^{\rm CIB}_{\rm lens}$&1.54$\pm$0.33 (4.7$\sigma$)&0.60$\pm$0.41 (1.5$\sigma$)&1.94$\pm$0.72 (2.7$\sigma$)&2.28$\pm$0.56 (4.0$\sigma$)\\
\hline
$\tilde\xi_{\rm NL}$&0.89$\pm$0.03&0.95$\pm$0.03&1.53$\pm$0.10&1.54$\pm$0.10\\
\hline
$\tilde\xi_{\rm NL}\bar I$[MJy/sr]&0.51$\pm$0.02&0.55$\pm$0.02&0.57$\pm$0.04&0.57$\pm$0.04\\
\hline
\end{tabular}}\label{ds}
\end{center}
\end{table*}

\section{Data analysis and results}
\label{data}

In this section we describe the data sets used in this work. We consider two sets of CMB lensing maps---a minimum-variance convergence map from the Planck public release 2 and the one from our own calculations using the Planck SMICA and SEVEM maps~\cite{2016A&A...594A...9P}. The component-separated CIB maps\footnotemark[3]\footnotetext[3]{Component-separated maps ``\rm{COM\_CompMap\_CIB-GNILC-F545\_2048\_R2.00.fits}'' and ``\rm{COM\_CompMap\_CIB-GNILC-F857\_2048\_R2.00.fits}'' are taken from Planck public release 2 \url{https://irsa.ipac.caltech.edu/data/Planck/release_2/all-sky-maps/}.} at Planck frequencies 857 and 545 GHz are used both to construct CIB lensing maps and as tracers of large-scale structure. To reduce any Galactic dust contamination, we mask out most of the low latitude regions, leaving only 17\% sky coverage for Planck CIB data as shown in Fig. \ref{mask}. The WISE galaxy count map is generated from the AllWISE source catalog\footnotemark[4]\footnotetext[4]{The all-sky WISE catalog is archived at \url{http://wise2.ipac.caltech.edu/docs/release/allsky/expsup/sec2_2.html}.} which contains about 0.6 billion objects detected at 3.4, 4.6, 12 and 22 $\mu m$ and the same data-cut criteria are applied to the map-making as listed in~\cite{2014A&A...571A..17P}. The resulting map of the galaxy number counts has a sky fraction of $0.27$. 

A suite of full-sky simulations described in Sec. \ref{sim} is made for the fluctuations of the CIB, galaxy density contrast and CMB lensing potential. Especially, the CIB realizations are created at three different levels of intrinsic non-Gaussianity with $\xinl=$ 0, 1, and 2, for two sets of maps---``unlensed+non-Gaussian'' and ``lensed+non-Gaussian''. Noise maps and instrumental specifications described in Sec. \ref{sim} are all taken into account for these simulations. The mock data are synthesized from these steps, and are processed through the procedures outlined in Sec. \ref{sim}.

The overall amplitudes of the measured CIB power spectra can match the theoretical predictions but there are still small perturbations in $\ell$ space. A correction which is a ratio of the measured CIB power spectrum to the theoretical prediction is precalculated at each frequency and applied to the simulated CIB maps $T$ when they are cross correlated with the reconstructed CIB lensing potentials from the simulations. The procedure outlined in Eqs. (\ref{rawest}--\ref{like}) is used to measure the bispectra. Both the parameters $A^{\rm CIB}_{\rm lens}$ and $\xinl$ are then sampled from the posterior distribution functions generated by Markov chain Monte Carlo (MCMC). Two sets of best-fit parameters are first derived from the MCMC chains for the single-frequency data that combine the three types of bispectra at 857 and 545 GHz. For each cross-frequency bispectrum $\langle T^{\alpha}(\xinl)T^{\alpha}(\xinl)\cdot T^{\beta}(\xinl')\rangle$, the non-Gaussian parameters $\xinl$ and $\xinl'$ are frequency dependent as reasons explained below, so both parameters should be varied for this case. Instead, we only fit for one parameter $\xinl$ and determine the value of $\xinl'$ from the ratio $\xinl'/\xinl$, which is set by the constrained values from the single-frequency bispectra. 

Longer CIB wavelengths trace star-forming galaxies at higher redshifts, so the level of the intrinsic non-Gaussianity that is projected onto a 2D map is lower than shorter CIB wavelengths. Thus, the intrinsic non-Gaussianity should be frequency dependent, in agreement with the measured intrinsic bispectra in Figs. \ref{datakt} and \ref{datakgp}. The CIB lensing is tightly related to the CIB source population and can also be affected by other cosmological parameters. Therefore, the best CIB model at each frequency can be fitted from a joint measurement including the CIB power spectrum and the three types of lensing bispectra. In this work, our focus is to perform a proof-of-concept study with the newly established technique, and defer a detailed investigation of astrophysical parameters and constraints to future work. The lensing amplitudes are within a factor of two of the nominal theoretical prediction as seen from Table \ref{ds}. We do not believe that this is a problem given the uncertainty in the prediction. The uncertain redshift distribution of the contributions to the CIB affects both the source and the lens. More positively, future measurements of this signal will be a sensitive probe of the CIB redshift distribution.

From the measurements shown in Figs. \ref{datakt} and \ref{datakgp}, we confirm that the intrinsic bispectra dominate the cross correlations $\langle\phi T\rangle$, $\langle\phi g\rangle$, and $\langle\phi\phi^{\rm CMB}\rangle$, whereas the lensing-induced bispectra are only subdominant. We subtract the intrinsic bispectra from the measured ones and obtain three types of bispectrum residuals, which are shown in Figs. \ref{datakt} and \ref{datakgp} with blue data points. The detection significance is calculated when each bispectrum residual is compared with the theoretical CIB lensing bispectrum predicted by the fiducial model in Sec. \ref{theory}. Detailed measurements for four different combinations are summarized in Table \ref{ds}. Combing all the measured bispectrum residuals, we detect an excess that is consistent with gravitational lensing in the secondary fluctuations of the cosmic infrared background at $>4\sigma$.

In addition to the validations for the calculations of the intrinsic-bispectrum outlined in Sec. \ref{validation}, other cross checks are also performed. The Galactic dust contamination is already minimized by the data cuts applied to the CIB data, including the conservative mask (high latitude regions) and a high-$\ell$ range ($1024<\ell<2048$) selection. Masking apodization is taken into account and an apodized mask is made by smoothing the 17\% mask with $\theta=$ 1 degree. Repeating the analysis with the apodized mask, we find that the overall lensing amplitude is only shifted by 5\%. To check the impact of foreground contaminants, we propagate a non-Gaussian dust template to the CIB maps and construct dust-contaminated CIB lensing maps. Calculations of the three types of cross correlations are repeated and the dust contributions are found to be negligible. To estimate possible dust residuals in the CMB lensing map, we replace the minimum-variance lensing map by either the ones derived from the Planck component-separated maps SMICA and SEVEM or the one reconstructed from polarization data only from 2018 Planck release. Both $\langle\phi\phi^{\rm CMB}\rangle$ cross-power spectra are consistent with the nominal case with the minimum-variance lensing map. Therefore, no foreground contaminants are detected from these tests.\\

\section{Conclusions}
\label{con}

In this work we apply a standard lensing-reconstruction technique from CMB data analysis to CIB fluctuations. It is found from numerical simulations that the intrinsic non-Gaussianity is the dominant non-Gaussian source among CIB fluctuations, making the detection of the gravitational lensing effects difficult. We cross correlate CIB lensing reconstructions with multiple tracers including the CIB itself. We propose a local non-Gaussian model for the CIB fluctuations and implement it in the estimation of the intrinsic cross correlations. Compared with sophisticated numerical simulations, it is found that the local non-Gaussian model can produce sufficiently accurate intrinsic bispectra. A cross-correlation technique that decouples lensing signals from the raw bispectra is thus established and verified by numerical simulations. Furthermore, it is applied to the Planck 857 and 545 GHz data. From the cross correlations between CIB lensing reconstructions and the tracers, an excess, which is consistent with CIB lensing effects, is detected at $>4\sigma$. Foregrounds and systematic effects are checked for the measurements and no significant contaminants are found. 

In the future, this technique will be further validated on more sophisticated numerical simulations of CIB fluctuations. CIB lensing maps will be reconstructed from lower frequency data such as the 280 GHz map of the Simons Observatory~\cite{2019JCAP...02..056A} and higher resolution data such as the Herschel Multi-tiered Extra-galactic survey~\cite{2010A&A...518L..21O}. High signal-to-noise CMB lensing maps reconstructed from polarization data of the South Pole Telescope~\cite{2017ApJ...849..124O}, Atacama Cosmology Telescope~\cite{2017PhRvD..95l3529S} and Simons Observatory, in conjunction with other galaxy samples, will be better tracers for more significant detection of CIB lensing effects.

\section{Acknowledgments}
We thank Dr. Mathieu Remazeilles for the ancillary data products pertaining to the component-separated CIB maps. We also thank Wayne Hu, Christopher Hirata and David Spergel for helpful discussions. This research is supported by the Brand and Monica Fortner Chair. We also acknowledge the use of the \hp~\cite{hp} package. This research used resources of the National Energy Research Scientific Computing Center (NERSC), a U.S. Department of Energy Office of Science User Facility operated under Contract No. DE-AC02-05CH11231. The sky simulations used in this paper were developed by the WebSky Extragalactic CMB Mocks team, with the continuous support of the Canadian Institute for Theoretical Astrophysics (CITA), the Canadian Institute for Advanced Research (CIFAR), and the Natural Sciences and Engineering Council of Canada (NSERC), and were generated on the Niagara supercomputer at the SciNet HPC Consortium. SciNet is funded by: the Canada Foundation for Innovation under the auspices of Compute Canada; the Government of Ontario; Ontario Research Fund - Research Excellence; and the University of Toronto.

\bibliography{ciblensing}

\end{document}